\documentclass[12pt]{article}
\usepackage{amsfonts,amsmath,amssymb}
\usepackage{theorem}
\usepackage{graphics,epsfig}
\numberwithin{equation}{section}

\newtheorem{definition}{Definition}[section]

\newtheorem{proposition}[definition]{Proposition}

\theorembodyfont{\rmfamily}

\def\cA{{\cal A}}

                    \def\cR{{\cal R}}

                    \def\mS{{\mathfrak S}}
\def\mW{{\mathfrak W}}

\newcommand{\CC}{{\mathbb C}}
\newcommand{\II}{{\mathbb I}}

\newcommand{\RR}{{\mathbb R}}
\def\lddots{\mathinner{\mkern1mu\raise1pt\hbox{.}\mkern2mu
\raise4pt\hbox{.}\mkern2mu\raise7pt\vbox{\kern7pt\hbox{.}}\mkern1mu}}

\newcommand{\ie}{{\it i.e.}\ }

\def\qmbox#1{\qquad\mbox{#1}\quad}

\begin{document}

\pagestyle{empty} \setcounter{page}{0}


\strut\hfill{}

\vspace{0.5in}

\begin{center}

{\Large \textsf{Exact results for the one-dimensional many-body problem  \\
with contact interaction: Including a tunable impurity}}

\vspace{10mm}

{\large V.Caudrelier $^a$\footnote{vc502@york.ac.uk} and
N.Cramp\'e $^b$\footnote{crampe@sissa.it}}

\vspace{10mm}

\emph{$^a$ Laboratoire d'Annecy-le-Vieux de Physique
Th{\'e}orique}

\emph{LAPTH, CNRS, UMR 5108, Universit{\'e} de Savoie}

\emph{B.P. 110, F-74941 Annecy-le-Vieux Cedex, France}

\vspace{7mm}
\emph{$^b$ Department of Mathematics\\
 University of
York\\
 Heslington York\\
  YO10 5DD, United Kingdom}

\end{center}

\vfill \vfill

\begin{abstract}
The one-dimensional problem of $N$ particles with contact
interaction in the presence of a tunable transmitting and
reflecting impurity is investigated along the lines of the
coordinate Bethe ansatz. As a result, the system is shown to be
exactly solvable by determining the eigenfunctions and the energy
spectrum. The latter is given by the solutions of the Bethe ansatz
equations which we establish for different boundary conditions in
the presence of the impurity. These impurity Bethe equations
contain as special cases well-known Bethe equations for systems on
the half-line. We briefly study them on their own through the
toy-examples of one and two particles. It turns out that the
impurity can be tuned to lift degeneracies in the energies and can
create bound states when it is sufficiently attractive. The
example of an impurity sitting at the center of a box and breaking
parity invariance shows that such an impurity can be used to
confine asymmetrically a stationary state. This could have
interesting applications in condensed matter physics.
\end{abstract}

\vfill MSC numbers: 82B23, 81R12, 70H06

 \vfill

\rightline{LAPTH-1083/05} \rightline{cond-mat/0501110}

\baselineskip=16pt

\newpage
\pagestyle{plain}

\section*{Introduction}

Forty years ago, E. Lieb and W. Liniger published their seminal
paper presenting exact results for the one-dimensional repulsive
Bose gas \cite{LL}, extending the previous investigation for
hard-core bosons \cite{girardeau}. This was completed in
\cite{mcguire} for the attractive interaction. It is remarkable that
this purely theoretical work finds a huge amount of applications
nowadays with the advent of optical lattices. The latter allow to
produce quasi one-dimensional environment where the quantum
behaviour of ultracold atoms can be probed experimentally
\cite{Paredes}. The main ingredient used in \cite{LL} is the
celebrated Bethe ansatz \cite{bethe} for the wavefunction. In
essence, this ansatz assumes an expansion of the wavefunction on
plane waves and the coefficients are determined so as to take the
interactions into account. Then, the energy spectrum is given by the
solution of the Bethe ansatz equations. Soon after, M. Gaudin
\cite{gaud} and then C. N. Yang \cite{yang} generalized the results
for particles with different type of statistics by considering a
wavefunction in different irreducible representation of the
permutation group. In particular, the investigation of C. N. Yang
relied on the now famous Yang-Baxter equation \cite{baxter}.
Finally, M. Gaudin studied also the analog of the system of
\cite{LL} when the bosons are enclosed in a box \cite{gaudin}. In
particular, he introduced a slightly more general Hamiltonian than
the contact interaction Hamiltonian of \cite{LL} depending on two
different coupling constants. The latter was recovered recently in
\cite{CC}, for particles with arbitrary spin, as a limit of a long
range interacting Hamiltonian of Sutherland type \cite{Sutherland}
for which integrability was proved. It was also shown that the
symmetry of this system is the reflection algebra symmetry
\cite{che,skly}. This motivates the interpretation of the
Hamiltonian considered by Gaudin as describing particles on the
half-line, or equivalently in the presence of a purely reflecting
impurity.

Let us stress that the many-body Hamiltonian of \cite{LL} is the
restriction to the $N$-particle Fock space of the well-known
nonlinear Schr\"odinger (NLS) Hamiltonian (see e.g. \cite{gutkin}
for a review). The NLS model is one of most studied examples of
integrable field theory for which a huge amount of exact results is
known. In the same way, the Hamiltonian of \cite{gaudin} is the
counterpart of the NLS model on the half-line whose symmetry is
given by the reflection algebra \cite{mintchev}, showing the
consistency of the approach of \cite{CC}. In \cite{mintchev}, the
concept of boundary algebra \cite{mintchev2} was crucial to
establish all the properties of NLS on the half-line as an
integrable system.

The question of a reflecting \textit{and} transmitting impurity in
integrable systems appeared naturally as a generalization of a
purely reflecting boundary. The first approach in this context was
done in \cite{DMS} where a set of reflection-transmission equations
was derived. Later in \cite{castro}, it was proved that non-trivial
solutions for the two-body scattering matrix do not exist if we
require a non-vanishing reflection and transmission coefficients for
the impurity. More recently, a new framework was introduced to
handle reflecting \textit{and} transmitting impurities in integrable
systems \cite{MRS} and was shown to be more general \cite{CMRS}. It
was successfully applied to the NLS model with impurity
\cite{CMR,CR} providing the first non-trivial integrable system with
impurity.

Consequently, it seemed natural to us to consider the many-body
analog of NLS with impurity and to investigate it along the lines
of \cite{LL} and \cite{gaudin}. Just like the system without
impurity, it may be of particular interest for current experiments
in condensed matter physics.

After presenting the problem in Section 1 together with some
notations to describe it, we show in Section 2 that it is exactly
solvable thanks to an appropriate Bethe ansatz for the
$N$-particle wavefunction. In Section 3, the full use of the Bethe
ansatz combined with the physical requirement of a finite size
system allows to establish the Bethe ansatz equations in the
presence of an impurity. This, in turn, is well-known to determine
the energy spectrum. Section 4 is devoted to specific examples.
First, we show that our setup reproduces the results of
\cite{gaudin} as a special case. Then, we use the one and
two-particle cases as toy examples to illustrate the effects of
the impurity on the energy levels and on the parity symmetry.
Finally, in Section 5, we present our conclusions for this work
and give an outlook of future investigations.

\section{The nature of the problem}\label{nature}

\subsection{Combining two systems}

In this paper, we study a one-dimensional system of $N$ particles
interacting through a repulsive $\delta$ potential in the presence
of an impurity sitting at the origin and described by a point-like
external potential. This problem is the combination of the
interacting system studied in \cite{LL,yang} and the free problem
in the presence of a point-like potential, see e.g. \cite{tsutsui}
and \cite{kurasov}.

Each of these problems has a well-defined translation in terms of
a partial differential equation problem together with boundary
conditions for the wavefunction. For example, let us denote by
$\varphi(x_1,\ldots,x_N)$ the $N$-particle wavefunction for a gas
with a repulsive $\delta$ interaction of coupling constant $g> 0$.
Then, following \cite{LL}, $\varphi$ is solution of the free
problem for the energy $E$
\begin{eqnarray}
\label{eigen} -\sum_{i=1}^N
\partial^2_{x_i}~\varphi(x_1,\ldots,x_N)&=&E~\varphi(x_1,\ldots,x_N)\,
,
\end{eqnarray}
with the additional requirement of continuity and jump in the
derivative at each hyperplane $x_j=x_k$, $j\neq k$
\begin{eqnarray}
\label{cont}
\varphi(x_1,\ldots,x_N)|_{~x_j=x_k^+}&=&\varphi(x_1,\ldots,x_N)|_{~x_j=x_k^-}\\
\label{jump}
(\partial_{x_j}-\partial_{x_k})~\varphi(x_1,\ldots,x_N)|_{~x_j=x_k^+}&=&[
(\partial_{x_j}-\partial_{x_k})+2g]\varphi(x_1,\ldots,x_N)|_{~x_j=x_k^-}
\end{eqnarray}
Now, in \cite{tsutsui}, the second problem is presented for the
one-particle wavefunction $\varphi(x)$, $x\neq 0$ using a unitary
matrix $U\in U(2)$ characterizing the impurity\footnote{in
\cite{tsutsui}, there is a length scale $L_0$ which is shown to be
an irrelevant parameter. We set it to $1$ in this paper.}:
\begin{equation}
\lim_{x\to 0^+}\left((U-\II)\Phi(x)+i(U+\II)\Phi'(x)\right)=0\, ,
\end{equation}
where
\begin{equation}
\Phi(x)= \left(\begin{array}{c}
          \varphi(x) \\
          \varphi(-x) \\
        \end{array}\right)~~,~~\Phi'(x)= \left(\begin{array}{c}
          \varphi'(x) \\
          -\varphi'(-x) \\
        \end{array}\right)~~,~~x>0\, ,
\end{equation}
$\varphi'(x)=d/dx ~\varphi(x)$ and $\II$ is the $2\times 2$ unit
matrix. The matrix $U$ can be parametrized as follows
\begin{equation}
\label{def_U}
U=e^{i\xi}\left(\begin{array}{cc}
                        \mu & \nu \\
                         -\nu^* & \mu^* \\
                       \end{array}\right)~~,~~\xi\in[0,\pi),~\mu,\nu\in\CC~~\text{such
                       that}~~|\mu|^2+|\nu|^2=1 \,
                       .
\end{equation}
The symbol $^*$ stands for complex conjugation. Mathematically,
this problem corresponds to all the possible self-adjoint
extensions of the free Hamiltonian when the point $x=0$ is removed
from the line.

As announced in the introduction, the goal of this paper is to
present and solve the quantum $N$-body problem combining these two
models. Physically speaking, we address the problem of a
one-dimensional gas of interacting particles in the presence of a
tunable impurity in the sense that the parameters in (\ref{def_U})
are free and can therefore be used to model different impurities
with different coupling constants.

\subsection{Notations and definitions}
{}From the mathematical point of view, the lesson we learn from
\cite{LL,yang} is the crucial role played by the permutation group
$\mS_N$ of $N!$ elements. It consists of $N$ generators: the
identity $Id$ and $N-1$ elements $T_1,\ldots,T_{N-1}$ satisfying
\begin{eqnarray}
\label{TT} &&T_j\,T_j=Id
\qmbox{,}T_{j}T_{\ell}=T_{\ell}T_{j} \qmbox{for} |j-\ell|>1\qmbox{,}\\
\label{TTT}
&&T_{j}T_{j+1}T_{j}=T_{j+1}T_{j}T_{j+1}\qmbox{.}
\end{eqnarray}
In particular, the last relation gives rise to the famous
Yang-Baxter equation \cite{yang,baxter}. For convenience, we
denote a general transposition of $\mS_N$ by $T_{ij}$, $i< j$,
given by
\begin{equation}
\label{generalT} T_{ij}=T_{j-1}\ldots T_{i+1}T_iT_{i+1}\ldots
T_{j-1}
\end{equation}

Then, in \cite{gaudin}, the role of the so-called reflection group
was emphasized and in \cite{mintchev2}, the Weyl group $\mW_N$
associated to the Lie algebra $B_N$ replaced the permutation group
in the construction of a Fock space for systems on the half-line.
Let us note that the same group proved to be fundamental in the
constructions of \cite{langmann} corresponding to an interacting
gas on the half-line where the usual $\delta$ interaction was
replaced by another contact interaction, the so-called $\delta'$
interaction. $\mW_N$ contains $2^NN!$ elements generated by $Id$,
$T_1,\ldots,T_{N-1}$ and $R_1$ satisfying (\ref{TT}), (\ref{TTT})
and
\begin{eqnarray}
&&R_1\,R_1=Id\qmbox{,}\\
&&R_1T_{1}R_1T_{1}=T_{1}R_1T_{1}R_1\qmbox{,}\\
&&R_1T_{j}=T_{j}R_1\qmbox{for} j>1\, .
\end{eqnarray}
Let us define also $R_j$, $j=2,\ldots,N$ as
\begin{equation}
R_j=T_{j-1}\ldots T_1R_1 T_1\ldots T_{j-1}
\end{equation}
Remarkably enough, the same group appears in the construction of
Fock space representations for systems with an impurity in the
context of RT algebras \cite{MRS}. One may wonder how the same
structure can account for systems on the half-line (\ie with
purely reflecting impurity) and also for systems on the whole line
with a reflecting and transmitting impurity. The essential point
is the choice of representation. Typically, for a system on the
half-line involving particles with $n$ internal degrees of
freedom, $n$-dimensional representations of $\mW_N$ are used. It
was realized in \cite{CMR,CR} that the same problem on the whole
line with a reflecting \textit{and} transmitting impurity requires
$2n$-dimensional representations of $\mW_N$. The interpretation of
this fact is that the impurity naturally defines two half-lines
which are physically inequivalent. Thus, in addition to the $n$
degrees of freedom of the internal symmetry, each particle carries
two degrees of freedom $+$ or $-$ labelling the side of the
impurity. These correspond to the two components of the
wavefunction $\Phi$ in (\ref{Phi}).

\section{Exact solvability of the model}

For pedagogical reasons, we present first the one and two-particle
cases in detail before turning to the study of the $N$-particle
problem in its full generality. We refer the experienced reader
directly to section \ref{Nparticles}.

\subsection{One particle}

For $x\in \RR\setminus\{0\}$, the one-particle wavefunction is
taken as follows
\begin{eqnarray}
\varphi(x)=
\begin{cases}
\varphi^+(x) & x>0\\
\varphi^-(x) & x<0
\end {cases}
\end{eqnarray}
Note that no parity relation is assumed for $\varphi$, which is a
crucial point in our approach. We define for $x>0$
\begin{eqnarray}
\label{Phi} \Phi(x)= \left(\begin{array}{c}
          \varphi^+(x) \\
          \varphi^-(-x) \\
        \end{array}\right)
\end{eqnarray}
and following the previous paragraph, the boundary conditions at
$x=0$, which we will call in this paper the impurity conditions,
read
\begin{eqnarray}
\label{bc} (U-\II)\Phi(x)=-i(U+\II)\Phi'(x)\qmbox{for}
x\rightarrow 0^+\, ,
\end{eqnarray}
the matrix $U$ being given in (\ref{def_U}). Let us expand $\Phi$
on plane waves as follows
\begin{eqnarray}
\Phi(x)= \exp(ik x)\cA_{Id}+ \exp(-ik x)\cA_{R}
\end{eqnarray}
where $\cA_P=\left(\begin{array}{c}
          A_P^+ \\
          A_P^- \\
       \end{array}\right)$, $P=Id,R$.
These coefficients are constrained by condition (\ref{bc}). This
is essentially the celebrated \textit{Bethe ansatz} for one
particle and it is solution of equation (\ref{eigen}) with
$E=k^2$. Plugging back into (\ref{bc}), one gets,
\begin{eqnarray}
\cA_{R}=Z(-k)\cA_{Id}~~\text{and}~~\cA_{Id}=Z(k)\cA_{R}
\end{eqnarray}
where
\begin{eqnarray}
\label{Z} Z(k)=-[U-\II-k  (U+\II)]^{-1}~[U-\II+k  (U+\II)]
\end{eqnarray}
The consistency of the ansatz is ensured by $Z(k)Z(-k)=\II$ which
is readily seen to hold. The property $Z^\dagger(k)=Z(-k)$, where
$^\dagger$ stands for Hermitian conjugation, then leads to the
physical unitarity $Z^\dagger(k)Z(k)=\II$.

For completeness, let us make the connection with the other usual
setting of the problem. For $\nu\neq 0$, (\ref{bc}) is equivalent
to
\begin{equation}
\label{standard}
\left(\begin{array}{cc} \varphi(x) \\
\varphi'(x) \end{array}\right) = \alpha \left(\begin{array}{cc} a
&
b\\ c&d\end{array}\right) \left(\begin{array}{cc} \varphi(-x)  \\
\varphi'(-x) \end{array}\right) \, ,\text{for}~~x\to 0^+ \, ,
\end{equation}
where
\begin{equation}
\{a,...,d \in \RR,\, \alpha \in \CC\, :\, ad -bc = 1,\, {\alpha^*}
\alpha = 1 \} \, .
\end{equation}
This is the $SU(2)$ parametrization. Writing $\mu=\mu_R+i\mu_I$,
$\nu=\nu_R+i\nu_I$ with $\mu_R$, $\mu_I$, $\nu_R$,
$\nu_I$$\in\RR$, the relation between the two parametrizations is
\begin{equation}
\label{relation} \alpha=\frac{i\nu}{|\nu|}~,~a=\frac{\sin
\xi-\mu_I}{|\nu|}~,~b=-\frac{\cos \xi+\mu_R}{|\nu|}~,~c=\frac{\cos
\xi-\mu_R}{|\nu|}~,~d=\frac{\sin \xi+\mu_I}{|\nu|}\, .
\end{equation}
{}From this one finds
\begin{equation}
\label{formZ}
Z(k)=\left(\begin{array}{cc}
                        R^+(k) & T^+(k) \\
                        T^-(-k) & R^-(-k) \\
                       \end{array}\right)
\end{equation}
where
\begin{eqnarray}
R^+(k) = \frac{bk^2 + i(a-d)k + c}{bk^2 + i(a+d)k - c} \, , \qquad
T^+(k) = \frac{2i\alpha k}{bk^2 + i(a+d)k - c}\, ,
\\
R^-(k) = \frac{bk^2 + i(a-d)k + c}{bk^2 - i(a+d)k - c} \, , \qquad
T^-(k) = \frac{-2i{\alpha^*} k}{bk^2 - i(a+d)k - c}\, ,
\end{eqnarray}
are usually referred to as reflection and transmission
coefficients of the impurity. Of great importance is the
well-known associated basis of orthonormal eigenfunctions for
scattering states
\begin{eqnarray}
\psi_k^+(x) = \theta (-x) T^-(k) e^{ikx} + \theta (x)\left[
e^{ikx} + R^+(-k)e^{-ikx}\right] \, , \quad k<0\, , \\
\psi_k^-(x) = \theta (x) T^+(k) e^{ikx} + \theta (-x)\left[e^{ikx}
+ R^-(-k)e^{-ikx}\right] \, , \quad k>0\, ,
\end{eqnarray}
 which appears as a particular choice of the above setting,
 justifying the Bethe ansatz approach.
These eigenfunctions play a crucial role in the quantum field
theoretic version of this problem \ie the nonlinear Schr\"odinger
equation with impurity \cite{CMR}.

For $\nu=0$, (\ref{bc}) gives rise to the so-called separated
boundary conditions of the form
\begin{eqnarray}
\varphi^{+'}(0^+)=q^+~\varphi^+(0^+)~~,~~\varphi^{-'}(-0^+)=q^-~\varphi^-(-0^+)
\end{eqnarray}
with $q^+
$, $q^-$$\in\RR\cup\{\infty\}$ given by
$q^\pm=\mp\tan(\frac{\xi\pm\zeta}{2})$, $\zeta$ being the argument
of $\mu$.

\subsection{Two particles}

In the same spirit as before, for $x_1,x_2 \in \RR\setminus\{0\}$
and $x_1\neq x_2$, the two-particle wavefunction is taken to be
\begin{eqnarray}
\label{2particles}
\varphi(x_1,x_2)=
\begin{cases}
\varphi^{++}(x_1,x_2) & x_1>0,x_2>0\\
\varphi^{+-}(x_1,x_2) & x_1>0,x_2<0\\
\varphi^{-+}(x_1,x_2) & x_1<0,x_2>0\\
\varphi^{--}(x_1,x_2) & x_1<0,x_2<0
\end {cases}
\end{eqnarray}
Then, we define for $x_1,x_2>0$ and $x_1\neq x_2$
\begin{eqnarray}
\Phi(x_1,x_2)= \left(\begin{array}{c}
          \varphi^{++}(x_1,x_2) \\
          \varphi^{+-}(x_1,-x_2) \\
          \varphi^{-+}(-x_1,x_2) \\
          \varphi^{--}(-x_1,-x_2) \\
        \end{array}\right)
\end{eqnarray}
Now, we implement the fact that each particle can interact with
the impurity by imposing two impurity conditions
\begin{eqnarray}
&&[(U-\II) \otimes \II]\Phi(x_1,x_2)=
-i[(U+\II) \otimes \II]\partial_{x_1}\Phi(x_1,x_2)\qmbox{for} x_1\rightarrow 0^+\\
&&[\II \otimes (U-\II)]\Phi(x_1,x_2)= -i[\II \otimes
(U+\II)]\partial_{x_2}\Phi(x_1,x_2)\qmbox{for} x_2\rightarrow 0^+
\, .
\end{eqnarray}
The interaction in the bulk between the two particles through a
$\delta$ potential is implemented as follows
\begin{eqnarray}
\label{bulk1}
\left.\Phi(x_1,x_2)\right|_{~x_1=x_2^+}&=&\widetilde{T}_{1}\left.\Phi(x_1,x_2)\right|_{~x_1=x_2^-}\\
\label{bulk2}
(\partial_{x_1}-\partial_{x_2})\left.\Phi(x_1,x_2)\right|_{~x_1=x_2^+}
&=&\widetilde{T}_{1}\left[(\partial_{x_1}-\partial_{x_2})+2g\right]\Phi(x_1,x_2)|_{~x_1=x_2^-}
\end{eqnarray}
where $\widetilde{T}_{1}$ is the representation on $\CC^2\otimes
\CC^2$ of $T_1\in \mS_2$ given by
\begin{eqnarray}
\widetilde{T}_{1}= \left(\begin{array}{c c c c}
          1& 0 &0& 0 \\
           0& 0 &1& 0\\
           0& 1 &0& 0\\
           0& 0 &0& 1\\
        \end{array}\right)
\end{eqnarray}
Similarly, $\widetilde{Id}$ is the $4 \times 4$ unit matrix
representing $Id$.

The crucial and new point now is to formulate an ansatz for
$\Phi(x_1,x_2)$ and show that it solves the problem. For
$0<x_{Q1}<x_{Q2}$ with $Q\in \mS_2=\{Id,T_{1}\}$, we take
\begin{eqnarray}
\Phi_Q(x_1,x_2)=\sum_{P\in ~\mW_2}
\exp[i(k_{P1}x_{Q1}+k_{P2}x_{Q2})]~~ \widetilde{Q}\cA_P(Q)
\end{eqnarray}
where $\cA_P(Q)=\left(\begin{array}{c}
A_P^{++}(Q) \\
          A_P^{+-}(Q) \\
          A_P^{-+}(Q) \\
          A_P^{--}(Q)
        \end{array}\right)$ are the coefficients to determine. The energy is simply
        $E=k_1^2+k_2^2$.\\
The impurity conditions imply
\begin{eqnarray}
\cA_{PR_1}(Id)&=&[Z(-k_{P1})\otimes \II]~ \cA_{P}(Id)\\
\widetilde{T}_{1}~\cA_{PR_1}(T_{1})&=&[\II \otimes Z(-k_{P1})]~
\widetilde{T}_{1}~ \cA_{P}(T_{1})
\end{eqnarray}
which reduce to
\begin{eqnarray}
\label{echangeZ} \cA_{PR_1}(Q)=[Z(-k_{P1})\otimes \II]~
\cA_{P}(Q)\qmbox{with} Q\in \mS_2
\end{eqnarray}
using
\begin{eqnarray}
\widetilde{T}_{1}~[\II \otimes Z(k)]~\widetilde{T}_{1}=
Z(k)\otimes \II \, .
\end{eqnarray}
The matrix $Z$ is the one given in (\ref{Z}).

The bulk conditions (\ref{bulk1}) and (\ref{bulk2}) give
\begin{eqnarray}
\label{echangeY} \cA_{PT_1}(Q)=\frac{1}{k_{P1}-k_{P2}+i g} \left(
(k_{P1}-k_{P2}) \cA_{P}(QT_1) -i g\cA_{P}(Q)\right) \qmbox{with}
Q\in \mS_2
\end{eqnarray}
Introducing the eight-component vector
\begin{eqnarray}
\label{vector}
\cA_{P}
        =\left(\begin{array}{c}
\cA_{P}(Id)\\
\cA_{P}(T_1)\\
        \end{array}\right)
\end{eqnarray}
We can rewrite (\ref{echangeZ}) and (\ref{echangeY}) in a compact
form
\begin{eqnarray}
\label{PR}
\cA_{PR_1}=[\II \otimes Z(-k_{P1})\otimes \II]~ \cA_{P}\\
\label{PT}
\cA_{PT_1}=Y(k_{P1}-k_{P2})\cA_{P}
\end{eqnarray}
where
\begin{eqnarray}
Y(k)= \left(\begin{array}{c c c c |c c c c}
\frac{-i g}{k+i g}&&&&\frac{k}{k+i g}\\
&\frac{-i g}{k+i g}&&&&\frac{k}{k+i g}\\
&&\frac{-i g}{k+ i g}&&&&\frac{k}{k +i g}\\
&&&\frac{-i g}{k+i g}&&&&\frac{k}{k +i g}\\
\hline\frac{k}{k+i g}&&&&\frac{-i g}{k + i g}\\
&\frac{k}{k+i g}&&&&\frac{-i g}{k+ i g}\\
&&\frac{k}{k+ i g}&&&&\frac{-i g}{k + i g}\\
&&&\frac{k}{k+i g}&&&&\frac{-i g}{k + i g}
\end{array}\right)
\end{eqnarray}
Since the relations $R_1^2=Id$, $T_1^2=Id$ and
$R_{1}T_{1}R_{1}T_{1}=T_{1}R_{1}T_{1}R_{1}$ hold in $\mW_2$,
equations (\ref{PR}) and (\ref{PT}) require that $Y(k)$ and $Z(k)$
satisfy the consistency relations
\begin{equation}
Z(k)Z(-k)=\II~~,~~Y(k_1-k_2)Y(k_2-k_1)=\II\otimes \II\otimes \II
\end{equation}
and a generalization of the celebrated reflection equation
\cite{che,skly},
\begin{eqnarray*}
Y(u-v)[ \II \otimes Z(u)\otimes \II]Y(u+v)[ \II\otimes Z(v)\otimes
\II] =[ \II\otimes Z(v)\otimes \II] Y(u+v)[ \II\otimes Z(u)\otimes
\II]Y(u-v)
\end{eqnarray*}
The explicit form of $Y$ and $Z$ ensures the validity of these
equations.

It is a generalization in the sense that even in the scalar case
(particles with no internal degrees of freedom), our setup
produces a two-dimensional representation of $\mW_2$. This is the
first illustration of the general statement at the end of Section
\ref{nature}.

We conclude that the two-particle model is exactly solvable in the
sense that the eigenfunction can be consistently given starting
from a given $\cA_P$, say $\cA_{Id}$.

\subsection{N particles}\label{Nparticles}

Following the previous arguments, we present the general solution
of our problem for $N$ particles and prove its exact solvability.

For $x_1,\dots, x_N \in \RR\setminus\{0\}$ and $x_1,\dots,x_N$ 2
by 2 different, the natural generalization of (\ref{2particles})
for the wavefunction is
\begin{eqnarray}
\varphi(x_1,\dots,
x_N)=\varphi^{\epsilon_1\,\dots\,\epsilon_N}(x_1,\dots,
x_N)\qmbox{in the region}\epsilon_1x_1>0,~\dots,~\epsilon_N x_N>0
\end{eqnarray}
where $\epsilon_i=\pm$, $i=1,\ldots,N$. Then, for $x_1,\dots, x_N
>0$ and $x_1,\dots,x_N$ 2 by 2 different, we define
\begin{eqnarray}
\label{def_Phi} \Phi(x_1,\dots, x_N) =
\sum_{\epsilon_1,\dots,\epsilon_N=\pm}
\varphi^{\epsilon_1\,\dots\,\epsilon_N}(\epsilon_1x_1,\dots,
\epsilon_N x_N)~~e_{\epsilon_1} \otimes \dots \otimes
e_{\epsilon_N}
\end{eqnarray}
where $e_+=\left(\begin{array}{c}
            1 \\
             0 \\
           \end{array}\right)$
           and
           $e_-=\left(\begin{array}{c}
            0 \\
             1 \\
           \end{array}\right)$.
Here we stress that the original wavefunction $\varphi$ is defined
for both signs of the $x_j$'s and that the wavefunction $\Phi$
contains the same physical information but is defined for
$x_1,\dots, x_N >0$ only. The advantage of the latter is that it
allows to impose all the conditions on the wavefunction
(interactions between particles in the bulk and boundary
conditions at the impurity) in a very compact form as we shall
see.

Given a tensor product of spaces, $(\CC^2)^{\otimes N}$, we define
the action of a matrix $M\in End(\CC^2)$ on the $k$-th space by
\begin{equation}
\label{tensor} M^{[k]}=\underbrace{\II \otimes \dots \otimes
\II}_{k-1} \otimes M \otimes \underbrace{\II \otimes \dots \otimes
\II}_{N-k}
\end{equation}
Therefore, the impurity conditions are,
\begin{eqnarray}
\label{bound_condition} (U-\II)^{[k]}~\Phi(x_1,\dots, x_N)=
-i(U+\II)^{[k]}~\partial_{x_k}\Phi(x_1,\dots, x_N)\qmbox{for}
x_k\rightarrow 0^+~~,~~1\leqslant k \leqslant N
\end{eqnarray}
The natural generalization of the bulk conditions read, for $Q\in
\mS_N$ and $1\leqslant i \leqslant N-1$,
\begin{eqnarray}
\label{bulk_condition1} &&\hspace{-1cm}\Phi(x_1,\dots,
x_N)|_{~x_{Qi}=x_{Q(i+1)}^+}
=\widetilde{Q}~\widetilde{T_{i}}~\widetilde{Q}^{-1}~\Phi(x_1,\dots, x_N)|_{~x_{Qi}=x_{Q(i+1)}^-}\\
&&\hspace{-1cm}(\partial_{x_{Qi}}-\partial_{x_{Q(i+1)}})~\Phi(x_1,\dots,
x_N)|_{~x_{Qi}=x_{Q(i+1)}^+}\nonumber\\
&&\hspace{+0.5cm}
=\widetilde{Q}~\widetilde{T_{i}}~\widetilde{Q}^{-1}~\left[
(\partial_{x_{Qi}}-\partial_{x_{Q(i+1)}})
+2g\right]\Phi(x_1,\dots,
x_N)|_{~x_{Qi}=x_{Q(i+1)}^-}\label{bulk_condition2}
\end{eqnarray}
$\widetilde{Q}$ is the usual representation of the element $Q\in
\mS_N$ on $(\CC^2)^{\otimes N}$. Namely, denoting by $E_{ij}$,
$i,j=1,2$ the matrices with $1$ at position $(i,j)$ and $0$
elsewhere, one has
\begin{equation}
\widetilde{T}_j=\sum_{k,\ell=1}^2 \underbrace{\II \otimes \dots
\otimes \II}_{j-1} \otimes E_{k\ell}\otimes E_{\ell k} \otimes
\underbrace{\II \otimes \dots \otimes \II}_{N-j-1}\, .
\end{equation}
Then using $\widetilde{T_iT_j}=\widetilde{T}_i\widetilde{T}_j$ and
(\ref{generalT}), it is easy to get $\widetilde{Q}$ for any $Q\in
\mS_N$ since an arbitrary permutation can always be decomposed in
transpositions. At this stage, we have explicitly formulated the
$N$-body problem corresponding to the combination of the two
systems as described in Section \ref{nature}.

Let us make the ansatz for $\Phi$: in the region
$0<x_{Q1}<\dots<x_{QN}$, $Q\in \mS_N$, it is represented by
\begin{eqnarray}
\label{ansatz} \Phi_Q(x_1,\dots, x_N)= \sum_{P\in~ \mW_N}
\exp[i(k_{P1}x_{Q1}+\dots+k_{PN}x_{QN})]~~ \widetilde{Q} \cA_P(Q)
\end{eqnarray}
where
\begin{eqnarray}
\cA_P(Q)= \sum_{\epsilon_1,\dots,\epsilon_N=\pm}
A_P^{\epsilon_1\,\dots\,\epsilon_N}(Q)~e_{\epsilon_1}\otimes \dots
\otimes e_{\epsilon_N}
\end{eqnarray}
Again, the eigenvalue problem is simply solved by
$E=\sum_{i=1}^N~k_i^2$.\\
Inserting in (\ref{bound_condition}), one gets
\begin{eqnarray}
\label{relZ} \cA_{PR_1}(Q)=Z^{[1]}(-k_{P1})~ \cA_{P}(Q)
\end{eqnarray}
where $Z$ is given by (\ref{Z}). From relations
(\ref{bulk_condition1}), (\ref{bulk_condition2}), we get for
$1\leqslant j \leqslant N-1$
\begin{eqnarray}
\label{relY} \cA_{PT_j}(Q)=\frac{1}{k_{Pj}-k_{P(j+1)}+i g} \left(
(k_{Pj}-k_{P(j+1)}) \cA_{P}(QT_j) -i g\cA_{P}(Q)\right)
\end{eqnarray}
To get an analog of (\ref{vector}), we introduce an ordering on
$\mS_N$ by associating to each element $Q\in \mS_N$ an integer
$[Q]\in \{1,\dots,N!\}$ so that $Q$ be the $[Q]^{\text{th}}$
element of the ordering list. Next, we define
\begin{eqnarray}
\label{vectorN} \cA_P=\sum_{Q\in \mS_N} e_{[Q]}\otimes \cA_P(Q)
\end{eqnarray}
where $e_{[Q]}=\begin{array}{c l l} \left(\begin{array}{c}
                 0 \\
                 \vdots \\
                 0 \\
                 1 \\
                 0 \\
                 \vdots \\
                 0 \\
               \end{array}\right)
&\begin{array}{l}\left\}\begin{array}{c}
                  \\
                  \\
                  \\
                  \end{array}\right.
                  \\
                  \\
                  \left\}\begin{array}{c}
                  \\
                  \\
                  \\
                  \end{array}\right.\\
               \end{array}
&\begin{array}{l}
                  \\
                 {}\hspace{-1cm}[Q]-1 \\
                  \\
                  \\
                  \\
                 \hspace{-1cm}N!-[Q] \\
                  \\
               \end{array}
\end{array}~~\in \CC^{N!}$\; so that $\cA_P(Q)$ is just
$(\cA_P)_{[Q]}$.\\
Thus, the relations (\ref{relZ}) and (\ref{relY}) take the compact
form
\begin{equation}
\label{relR} \cA_{PR_1}=\mathcal{Z}_1(-k_{P1})~\cA_{P}
\end{equation}
 and for $1\leqslant j \leqslant N-1$
\begin{eqnarray}
\label{relT}
\cA_{PT_j}=Y_j(k_{Pj}-k_{P(j+1)})\cA_P
\end{eqnarray}
where the matrix elements of $\mathcal{Z}$ and $Y_j$ read (recall
that these matrix elements are matrices themselves acting on
$(\CC^2)^{\otimes N}$)
\begin{eqnarray}
\mathcal{Z}_1(k)_{[Q],[Q']}&=&Z^{[1]}(k) \delta_{[Q],[Q']}\\
Y_j(k)_{[Q],[Q']}&=&\frac{1}{k+i g} \left(~k~
\delta_{[QT_j],[Q']}-i g~\delta_{[Q],[Q']} \right)\II^{\otimes N}
\end{eqnarray}
Since our construction is based on $\mW_N$, the Bethe ansatz
solution is consistent if $\mathcal{Z}_1$ and $Y_j$ satisfy
\begin{eqnarray}
\label{unit} &&Y_{j}(k)Y_{j}(-k)=\II_{N!}\otimes \II^{\otimes
N}\qmbox{,} \mathcal{Z}_1(k)\mathcal{Z}_1(-k)=\II_{N!}\otimes
\II^{\otimes N}
\\
\label{YBE}
&&Y_{j}(k_1)Y_{j+1}(k_1+k_2)Y_{j}(k_2)=Y_{j+1}(k_2)Y_{j}(k_1+k_2)Y_{j+1}(k_1)
\\
\label{RE}
&&Y_{1}(k_1-k_2)\mathcal{Z}_1(k_1)Y_{1}(k_1+k_2)\mathcal{Z}_1(k_2)=
\mathcal{Z}_1(k_2)Y_{1}(k_1+k_2)\mathcal{Z}_1(k_1)Y_{1}(k_1-k_2)
\\
&&Y_{j}(k_1)Y_{\ell}(k_2)=Y_{\ell}(k_2)Y_{j}(k_1)\qmbox{for} |j-\ell|>1~,
\\
\label{ZZ}
&&\mathcal{Z}_1(k_1)Y_{j}(k_2)=Y_{j}(k_2)\mathcal{Z}_1(k_1)\qmbox{for}
j>1
\end{eqnarray}
where $\II_{N!}$ the $N! \times N!$ unit matrix. Relations
(\ref{unit}) are usually called unitarity conditions while
(\ref{YBE}) is the celebrated quantum Yang-Baxter equation
\cite{yang,baxter}. Relation (\ref{RE}) is again our generalized
reflection equation. One can check that these relations hold true
by direct computation (whatever the values of $g$ and $\xi$, $\mu$
and $\nu$ defined in (\ref{def_U})), finishing our argument about
the exact solvability of our $N$-particle system. Starting from
$\cA_{Id}$ and using (\ref{relR}) and (\ref{relT}) repeatedly, one
gets the eigenfunction.

\section{Bethe ansatz: spectrum in the presence of an impurity}

In the previous section, we showed that the energy problem reads
\begin{equation}
E=\sum_{i=1}^N~k_i^2
\end{equation}
where the $k$'s are the momenta of the particles. It is known that
the complete use of the Bethe ansatz entails that the $k$'s are
the solutions of the so-called Bethe ansatz equations. From these
equations, it is possible to get some insight in the energy
spectrum of the problem. The usual approach is to enclose the
system in a finite region of space. One can imagine two types of
conditions at the border of the finite region. In one dimension,
one can put the $N$ particles on a circle requiring periodic (or
even anti-periodic) condition. This was the choice made in
\cite{LL} where the properties on the whole line were subsequently
extracted through the so-called thermodynamic limit. An
alternative approach is to enclose the particles in a box
requiring the vanishing of the wave function on the walls of the
box. This was explore e.g. in \cite{gaudin}.

\subsection{Bethe ansatz equations for particles on a circle}

Let us imagine that the $N$ particles live on the interval
$[-L,L]$ centered for convenience on the impurity. In terms of the
original wavefunction $\varphi$, the periodic (resp.
anti-periodic) condition on the $\ell$-th particle, $1\leqslant
\ell \leqslant N$, reads,
\begin{eqnarray}
\label{perio1}
\varphi(x_1,\dots,x_{\ell-1},L,x_{\ell+1}\dots,x_{N})=
\theta~\varphi(x_1,\dots,x_{\ell-1},-L,x_{\ell+1}\dots,x_{N})\\
\label{perio2}
\varphi'(x_1,\dots,x_{\ell-1},L,x_{\ell+1}\dots,x_{N})=
\theta~\varphi'(x_1,\dots,x_{\ell-1},-L,x_{\ell+1}\dots,x_{N})\, ,
\end{eqnarray}
with $\theta=1~(\text{resp.}~\theta=-1)$. Introducing
\begin{equation}
\sigma=\theta\left(
 \begin{array}{cc}
   0 & 1 \\
   1 & 0 \\
        \end{array}
        \right)\, ,
\end{equation}
and using the tensor notations (\ref{tensor}), equations
(\ref{perio1}) and (\ref{perio2}) can be equivalently written in
terms of $\Phi$ as
\begin{eqnarray}
\label{L_condition1}
\Phi(x_1,\dots,x_{\ell-1},L,x_{\ell+1}\dots,x_{N})
=\sigma^{[\ell]}~
\Phi(x_1,\dots,x_{\ell-1},L,x_{\ell+1}\dots,x_{N})\\
\label{L_condition2}
\partial_{x_\ell}\Phi(x_1,\dots,x_{\ell-1},L,x_{\ell+1}\dots,x_{N}) =
-\sigma^{[\ell]}~
\partial_{x_\ell}\Phi(x_1,\dots,x_{\ell-1},L,x_{\ell+1}\dots,x_{N})
\end{eqnarray}
Invoking the Bethe ansatz solution (\ref{ansatz}) for some
$Q_\ell\in \mS_N$ such that $Q_\ell(N)=\ell$, one gets (noting
that $\widetilde{Q}_\ell~
\sigma^{[\ell]}=\sigma^{[N]}~\widetilde{Q}_\ell$)
\begin{eqnarray}
e^{ik_{PN}L}\cA_P(Q_\ell)+e^{-ik_{PN}L}\cA_{PR_N}(Q_\ell) =
\sigma^{[N]}~
\big({e^{ik_{PN}L}\cA_P(Q_\ell)+e^{-ik_{PN}L}\cA_{PR_N}(Q_\ell)}\big)\\
e^{ik_{PN}L}\cA_P(Q_\ell)-e^{-ik_{PN}L}\cA_{PR_N}(Q_\ell) =
-\sigma^{[N]}~
\big({e^{ik_{PN}L}\cA_P(Q_\ell)-e^{-ik_{PN}L}\cA_{PR_N}(Q_\ell)}\big)
\end{eqnarray}
This entails
\begin{equation}
e^{2ik_{PN}L}\cA_P(Q_\ell)-\sigma^{[N]}\cA_{PR_N}(Q_\ell)=0~~,~~\ell=1,\ldots,N\,
\end{equation}
that is, in terms of $\cA_P$ as defined in (\ref{vectorN})
\begin{equation}
\label{constraint}
e^{2ik_{PN}L}\cA_P-\Sigma\cA_{PR_N}=0\,,~~\Sigma=\II_{N!}\otimes
\sigma^{[N]}~ .
\end{equation}
This holds for any $P\in \mW_N$ yielding \textit{a priori} $2^N
N!$ different equations. In fact, let us show that we only need to
consider $N$ of them by proving that if (\ref{constraint}) holds
for $\cA_P$ then it holds for $\cA_{PT_j}$, $j=1,\ldots,N-2$,
$\cA_{PR_1}$ and $\cA_{PR_N}$. For $j=1,\ldots,N-2$
\begin{eqnarray}
\cA_{PT_j}&=&Y_j(k_{Pj}-k_{P(j+1)})\cA_P\\
&=&
e^{-2ik_{PN}L}Y_j(k_{Pj}-k_{P(j+1)})\Sigma\cA_{PR_N}\\
&=& e^{-2ik_{PN}L}\Sigma Y_j(k_{Pj}-k_{P(j+1)})\cA_{PR_N}\\
&=& e^{-2ik_{PN}L}\Sigma\cA_{PR_NT_j}\\
&=& e^{-2ik_{PN}L}\Sigma\cA_{PT_jR_N}
\end{eqnarray}
where we used $Y_j(k)\Sigma=\Sigma Y_j(k)$ and $R_NT_j=T_jR_N$.
The proof for the other two cases is similar and requires
$\mathcal{Z}_1(k)\Sigma=\Sigma \mathcal{Z}_1(k)$, $R_NR_1=R_1R_N$
and $\Sigma^2=\II_{N!}\otimes \II^{\otimes N}$. Since $T_j$,
$j=1,\ldots,N-2$ and $R_1$ are the generators of $W_{N-1}$ of
cardinal $2^{N-1}(N-1)!$, adding $R_N$ brings the number of
elements to $2^{N}(N-1)!$. Therefore, quotienting $\mW_N$ by this
set, we are left with $N$ different elements: $S_N=Id$ and
$S_j=T_j\ldots T_{N-1}$ for $j=1,\ldots,N-1$.

Now using (\ref{relR}) and (\ref{relT}) repeatedly, one has
\begin{eqnarray}
\label{relPR} \cA_{S_jR_N}&=&Y_{N-1}(-k_{N}-k_{j})\dots
Y_{1}(-k_{1}-k_{j})\nonumber\\
&&~~\times~~\mathcal{Z}_1(-k_{j}) Y_{1}(k_{1}-k_{j}) \dots
Y_{N-1}(k_{N}-k_{j}) \cA_{S_j}\, ,
\end{eqnarray}
and
\begin{equation}
\label{relS} \cA_{S_j}=Y_{N-1}(k_{j}-k_{N})\ldots
Y_{j}(k_{j}-k_{j+1})\cA_{Id}\, .
\end{equation}
Let us introduce the matrices $\cR_j$ for $j=1,\ldots,N$ as
\begin{eqnarray}
\label{def_Rj} \cR_j&=& Y_{j}(k_{j+1}-k_{j}) \ldots
Y_{N-1}(k_{N}-k_{j})~\Sigma ~Y_{N-1}(-k_{N}-k_{j})\dots
Y_{j}(-k_{j+1}-k_{j})\nonumber\\
&&~\times Y_{j-1}(-k_{j-1}-k_{j})\dots Y_{1}(-k_{1}-k_{j})
\mathcal{Z}_1(-k_{j}) Y_{1}(k_{1}-k_{j}) \dots
Y_{j-1}(k_{j-1}-k_{j})
\end{eqnarray}
Applying all these results in (\ref{constraint}), we are now in
position to state the main result of this paper:
\begin{proposition}
The wavefunction of our exactly solvable model is completely
determined for a given vector $\cA_{Id}$, using relations
(\ref{relR}) and (\ref{relT}) to find $\cA_P$ for any $P\in
\mW_N$. In turn, $\cA_{Id}$ is the common eigenvector of the
matrices $\cR_j$ with the eigenvalues $e^{2ik_{j}L}$ respectively,
$j=1,\ldots,N$:
\begin{eqnarray}
\label{det1} \cR_j~ \cA_{Id}=e^{2ik_j L}~ \cA_{Id}\, .
\end{eqnarray}
This entails in particular the following constraints
\begin{equation}
 \det \left[\cR_j-e^{2ik_{j}L}~\II_{N!}\otimes \II^{\otimes N}\right]=0~~,~~j=1,\ldots,N\, .
\end{equation}
These are the impurity Bethe ansatz equations constraining the
allowed values of the momenta of the particles. The presence of
$Z$ in $\cR_j$ accounts for the effect of the impurity on the
dynamics of the system while the matrices $Y_j$ contain the
interaction effects.
\end{proposition}
The proof goes as follows. First, relation (\ref{det1}) is a
direct consequence of (\ref{constraint}), (\ref{relPR}) and
(\ref{relS}). The fact that $\cA_{Id}$ is the common eigenvector
of all these matrices follows from
\begin{equation}
\cR_j~\cR_\ell=\cR_\ell~\cR_j~~,~~j,\ell=1,\ldots,N\, .
\end{equation}
This equality, albeit tedious to establish, holds thanks to
relations (\ref{unit})-(\ref{ZZ}) together with
\begin{equation}
\left[Y_j(k_1),Y_j(k_2)\right]=0~~,~~\text{for all}~k_1,k_2\,.
\end{equation}

\subsection{Fixing the statistics}

So far, we said nothing about the statistics of the particles
under considerations (on purpose). Indeed, our setup can be
accommodated along the lines of \cite{yang} to allow for arbitrary
statistics. Here, to get more insight when dealing with the Bethe
ansatz equations, let us choose the statistics of our model. For
bosons (resp. fermions), the wavefunction should be symmetric
(resp. antisymmetric) under the exchange of any two particles. In
terms of $\Phi$, this reads, for $1 \leqslant i < j \leqslant N$,
\begin{eqnarray}
\label{statistics}
\Phi(x_1,\dots,x_i,\dots,x_j,\dots,x_N)=\tau~
\widetilde T_{ij}~~ \Phi(x_1,\dots,x_j,\dots,x_i,\dots,x_N)
\end{eqnarray}
with $\tau=+1$ for bosons and $\tau=-1$ for fermions. In turn,
this yields an additional relation between the coefficients
$\cA_P(QT_i)$ and $\cA_P(Q)$:
\begin{eqnarray}
\cA_P(QT_i)=\tau\cA_P(Q)\, .
\end{eqnarray}
As a consequence, all the matrices $Y_{j}(u)$ become proportional
to the identity, the multiplication factor being $y^\tau(k)$ given
by
\begin{eqnarray}
\label{y_scalar}
y^\tau(k)=\frac{\tau k-i g}{k+i g}
\end{eqnarray}
When $\tau=-1$, we recover that the $\delta$ interaction is
trivial for spinless fermions. The $N$ impurity Bethe equations
are equivalent to
\begin{eqnarray}
\label{bethe_eq} &&\det\left[\prod_{m \neq j}y^\tau(k_{j}+k_{m})
y^\tau(k_{j}-k_{m})\exp(2ik_{j}L)- Z^{[1]}(-k_j) \sigma^{[N]}
\right]=0~~,~~1\leqslant j \leqslant N \,
\end{eqnarray}
Let us denote $z_1(k)$, $z_2(k)$ the eigenvalues of $Z(k)$ then,
 for $N\ge 2$, $Z^{[1]}(k) \sigma^{[N]}$ has four different
 eigenvalues $z_1(k)$, $-z_1(k)$, $z_2(k)$ and $-z_2(k)$, each of
 which is $2^{N-2}$-fold degenerate. For $N\ge 2$, the $N$ equations (\ref{bethe_eq})
 are in turn equivalent to
\begin{eqnarray}
\label{bethe_eqd} &&\exp(2ik_{j}L)~\prod_{m \neq
j}y^\tau(k_{j}+k_{m}) y^\tau(k_{j}-k_{m})=\lambda_j(k_j)
~~,~~1\leqslant j \leqslant N \,
\end{eqnarray}
where for each $j$, $\lambda_j(k_j)$ takes one of the four possible
values $z_1(-k_j)$, $-z_1(-k_j)$, $z_2(-k_j)$ or $-z_2(-k_j)$. Then,
to find the complete spectrum, we need to solve the set of $N$
equations (\ref{bethe_eqd}) for the $N$ unknowns $\{k_j\}$. Since,
for each equation, four choices of $\lambda_j$ are possible, there
are $4^N$ different sets of $N$ equations to solve.

For $N=1$, the equations read
\begin{eqnarray}
\exp(2ikL)= s_1(-k) \qmbox{or} \exp(2ikL)= s_2(-k)
\end{eqnarray}
where $s_1(k)$, $s_2(k)$ are the eigenvalues of $\sigma Z(k)$. In
the previous equations, we isolated on the left-hand side the
usual terms corresponding to the interaction and on the right-hand
side the new part arising from the presence of the reflecting and
transmitting impurity.

\subsection{Bethe ansatz equations for particles in a box}\label{box}

In this paragraph, instead of putting the particle on a circle, we
are going to let them live in a box $[-L,L]$. Thus, we impose the
following conditions for $1\leqslant \ell \leqslant N$
\begin{eqnarray}
\Phi(x_1,\dots,x_{\ell-1},L,x_{\ell+1}\dots,x_{N})=0 \, ,
\end{eqnarray}
and we follow the above analysis along the same lines. The linear
system of equations relating the $2^NN!$ coefficients in each
$\cA_{S_j}$, $j=1,\ldots,N$, take the form
\begin{eqnarray}
\cR_j\cA_{Id}=-e^{2ik_jL}\cA_{Id}
\end{eqnarray}
with $\cR_j$ as defined in (\ref{def_Rj}) replacing $\Sigma$ by
the identity here. In this case, the impurity Bethe equations read
\begin{equation}
 \det \left[\cR_j+\exp(2ik_{j}L)~\II_{N!}\otimes \II^{\otimes N}\right]=0~~,~~j=1,\ldots,N\, .
\end{equation}
These are the Bethe ansatz equations for our system with the
particular box conditions under consideration. If we fix the
statistics for bosons or fermions as in the previous paragraph,
these equations take the simpler form
\begin{eqnarray}
&&\exp(2ik_{j}L)~\prod_{m \neq j}y^\tau(k_{j}+k_{m})
y^\tau(k_{j}-k_{m})=\gamma_j(k_j) ~~,~~1\leqslant j \leqslant N \,
\end{eqnarray}
where for each $j$, $\gamma_j(k_j)$ takes two possible values :
$z_1(-k_j)$ or $z_2(-k_j)$, yielding $2^N$ different sets of $N$
equations. This setup will be used in the examples to illustrate the
case of a parity-breaking impurity.

\section{Detailed study of selected examples}

\subsection{Recovering previous results}

Let us show how to recover the historical results of \cite{gaudin}
which are the analog of the results of \cite{LL} when the
particles are confined in a box.

We recall briefly the setup of \cite{gaudin} and adapt ours to
reproduce it. M. Gaudin considered a gas of $N$ bosons with
$\delta$ interaction enclosed in a box of length $L$. The
symmetric wavefunction $\phi(x_1,\ldots,x_N)$ is required to
satisfy
\begin{eqnarray}
\label{cond_gaudin}
\phi(x_1=0,x_2,\ldots,x_N)=0\, ,\\
\label{cond_gaudin2}
\phi(x_1,x_2,\ldots,x_N=L)=0\, ,
\end{eqnarray}
in the region $0\le x_1 \le x_2 \le \ldots \le x_N \le L$.

Thus we must take (\ref{y_scalar}) with $\tau=1$. Then, the most
natural idea that comes to mind to recover this setup from ours is
to "fold" our system which lives on $[-L,L]$ and tune the
parameters of the impurity so as to make it a wall of the box at
the origin. This goes as follows: for $j=1,\ldots,N$, we require
\begin{equation}
\label{parity}
\varphi(x_1,\ldots,x_j,\ldots,x_N)=\varphi(x_1,\ldots,-x_j,\ldots,x_N)~~,~~0<x_j<L
\end{equation}
This global property has a direct consequence on $\Phi$ as defined
in (\ref{def_Phi})
\begin{eqnarray}
\Phi(x_1,\ldots,x_N)=\varphi^{+\ldots
~+}(x_1,\ldots,x_N)~\begin{array}{c l l}

                      \left(\begin{array}{c}
                 1 \\
                 \vdots \\
                  1 \\
                  \end{array}\right)
&\begin{array}{l}\left\}\begin{array}{c}
                  \\
                  \\
                  \\
                  \end{array}\right.
                  \end{array}
&\begin{array}{l}
                  \\
                 {}\hspace{-1cm}2^N \\
                  \\

                 \end{array}
\end{array}\, .
\end{eqnarray}
In other words, the representation is completely reducible and the
only relevant wavefunction is $\varphi^{+\ldots
~+}(x_1,\ldots,x_N)$ which we identify to $\phi(x_1,\ldots,x_N)$.
The reducibility of the problem will show up again consistently in
the rest of this paragraph (e.g. for $Z(k)$). It has to be related
to the fact that we break the chirality of the impurity when we
require (\ref{parity}), that is we restore the parity invariance
and the need for a two-dimensional representation disappears.

In this respect, we also expect (\ref{parity}) not to be
compatible with the impurity conditions (\ref{bound_condition}) in
general. In fact, it is easy to see that the coefficients $\mu$,
$\nu$ in (\ref{def_U}) must satisfy
\begin{equation}
\mu-\mu^*=0~~,~~\nu+\nu^*=0\, .
\end{equation}
When translated in terms of $\alpha, a,b,c,d$ in (\ref{relation}),
this is perfectly consistent with the well-known characterization
of a parity-invariant impurity \ie $a=d$ and  $\alpha^2=1$.

Now to reproduce (\ref{cond_gaudin}), one just has to choose
$\mu=-1$ and $\nu=0$. In particular, this gives $Z(k)=-\II$
showing again the reducibility of the problem to a scalar
representation where the impurity is to be seen as a purely
reflecting wall with reflection coefficient equal to $-1$.

Finally, taking $\theta=-1$ in (\ref{L_condition1}),
(\ref{L_condition2}) yields (\ref{cond_gaudin2}) with no condition
for the derivative as required. Again, the representation using
$\sigma$ is reducible and one can see that the relevant eigenvalue
for sigma is $-1$. Collecting all these settings, we end up with
the following Bethe equations
\begin{equation}
e^{2ik_jL}=\prod_{m\neq
j}\frac{k_m-k_j-ig}{k_m-k_j+ig}\frac{k_m+k_j-ig}{k_m+k_j+ig}~~,~~j=1,\ldots,N
\end{equation}
which are precisely those obtained by Gaudin in \cite{gaudin}.

\subsection{One and two particles with $\delta$ impurity}

As a first step towards the understanding of the properties of our
system, we pay special attention to the special cases of one and
two particles in the presence of the well-known $\delta$ impurity.
The one-particle case is presented as a reminder and already
displays the interesting features of degeneracies and bound
states. Then, the two-particle exhibits the new properties arising
from the impurity Bethe equations.

The $\delta$ impurity is characterized by
\begin{eqnarray}
U_\delta=-\exp(i\xi) \left(
\begin{array}{cc}
  \cos(\xi) & i\sin(\xi) \\
  i\sin(\xi) & \cos(\xi) \\
\end{array}
\right)~~ ,~~\xi\in[0,\pi)
\end{eqnarray}
and the impurity Bethe equations constraining the momentum of the
particle read
\begin{eqnarray}
\label{bethe_delta} \exp(2ikL)=\theta \qmbox{or}
\exp(2ikL)=\theta\, \frac{k\tan \xi +i}{k\tan \xi-i}
\end{eqnarray}
The first equation reproduces the usual integer quantization of
the momentum while the second shows how this quantization is
controlled by $\xi$.

\begin{figure}[htp]
\begin{minipage}[b]{8cm}
\epsfig{file=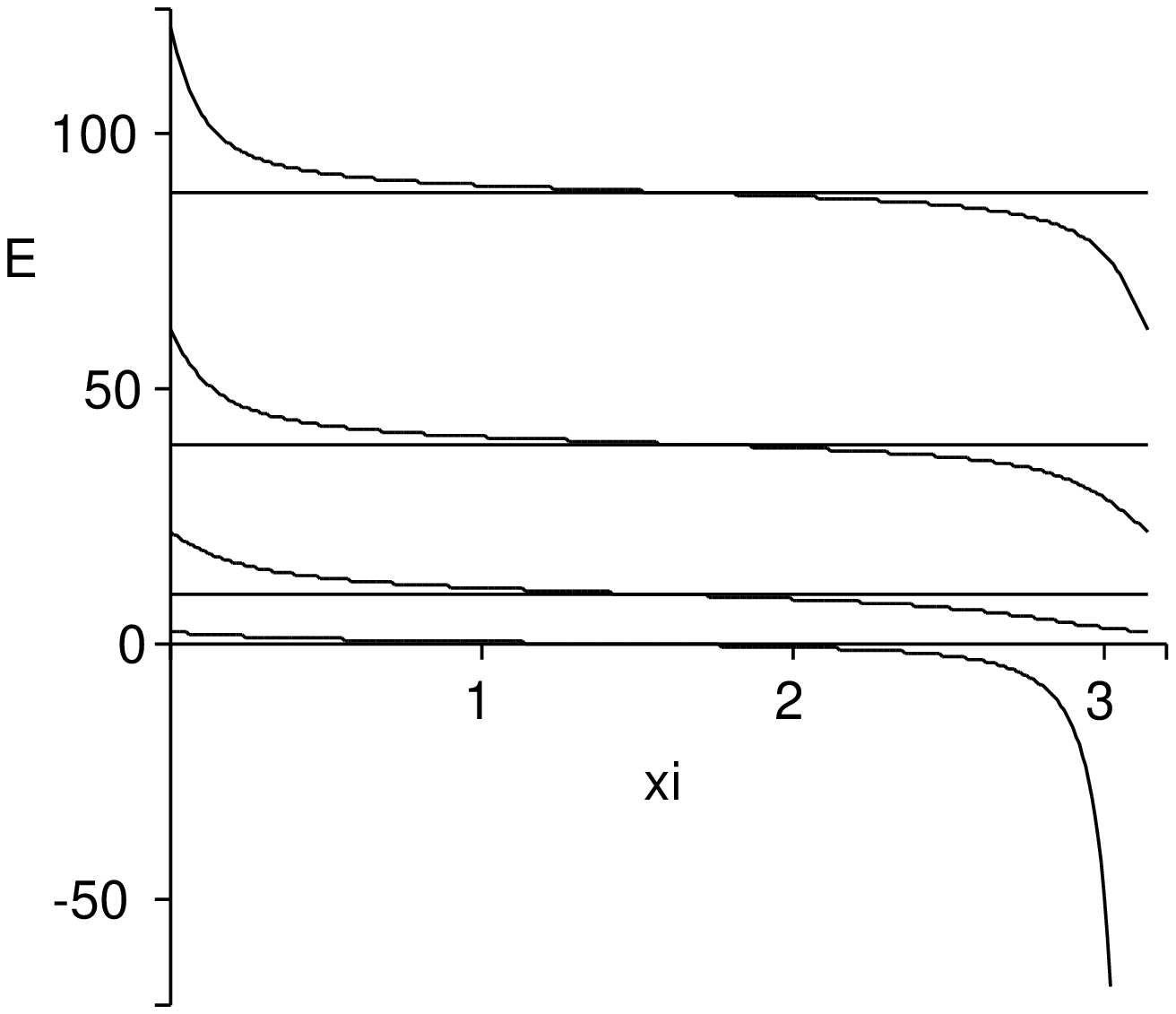,width=8cm}
\caption{\label{k_deltaP}Lowest energy level in terms of $\xi$ for
$\delta$ impurity and $\theta=1$}
\end{minipage}
\qquad
\begin{minipage}[b]{8cm}
\epsfig{file=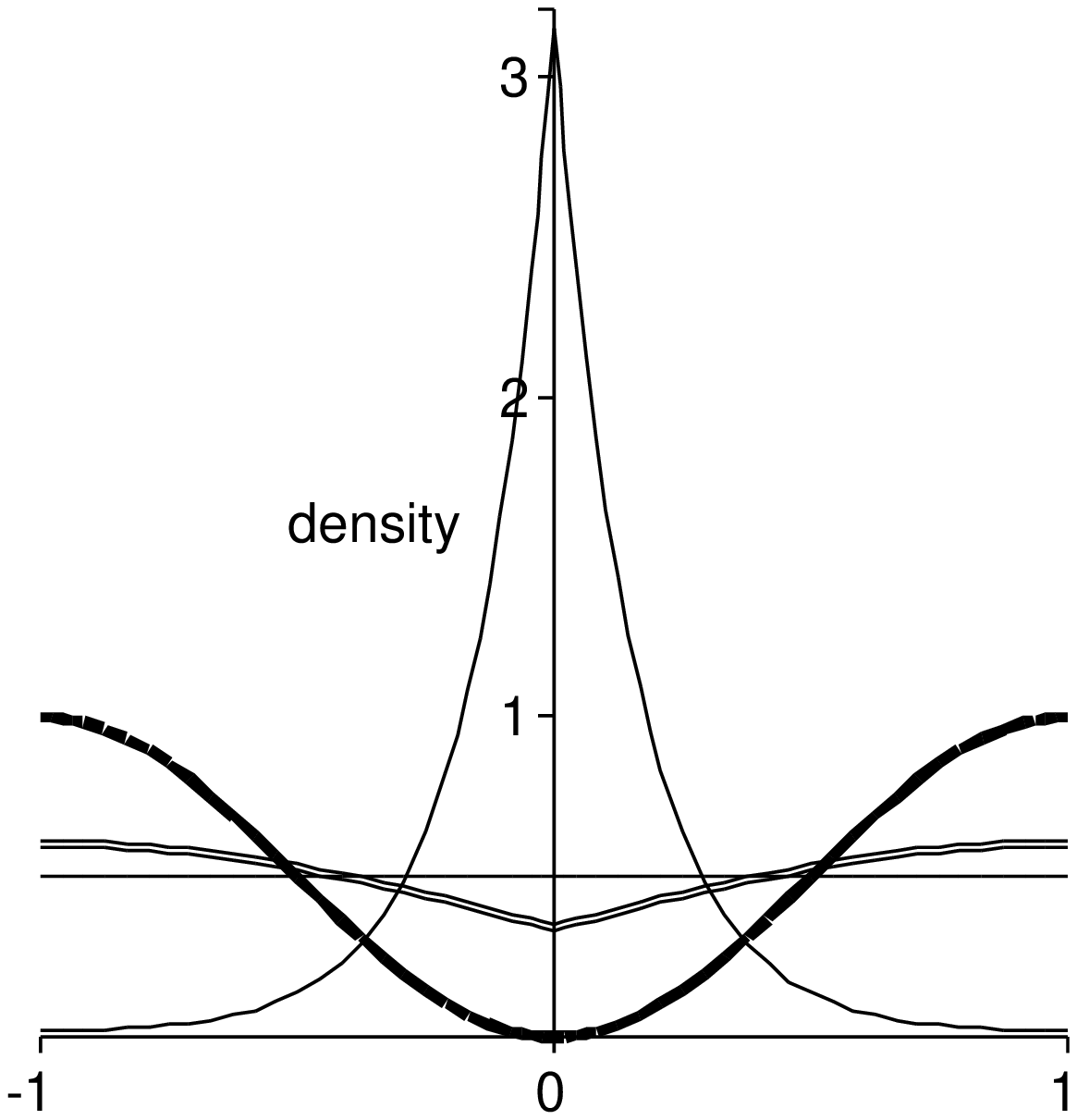,width=8cm}
\caption{\label{density_delta} Density (=$|\varphi(x)|^2$) for
various values of $\xi$.}
\end{minipage}
\end{figure}

Figure \ref{k_deltaP} shows the energy spectrum\footnote{all the
figures are plotted in units of $\hbar^2/2m$, for $g=1$ and for a
unit length, $L=1$, using Maple.} as a function of the tunable
impurity parameter $\xi$.

The constant energy levels correspond to the first equation and do
not depend on the impurity parameter $\xi$ as expected. The other
energy levels show the effect of the impurity on the spectrum. Of
special interest is the lowest energy level which exhibits a bound
state for $\xi>\pi/2$. This is consistent with the fact that the
coupling constant to the impurity $\eta$, given by
$\eta=1/\tan\xi$, becomes negative. We have also plotted in Figure
\ref{density_delta} the corresponding densities for different
regimes. The thick curve corresponds to $\xi=0$: the impurity is
completely reflecting and no transmission occurs. For $\xi=\pi/3$,
the double-solid curve shows reflection and transmission for a
repulsive impurity. The constant curve for $\xi=\pi/2$ is very
special since for this value of $\xi$, the impurity becomes
trivial in the sense that the reflection vanishes and the
transmission is just $1$. This corresponds to the zero energy
state. Finally, the thin curve represents the profile for
$\xi=11\pi/12$: this is the bound state whose profile gets sharper
and sharper as $\xi\to\pi$ (infinitely attractive impurity).\\

\begin{figure}[htp]
\begin{center}
 \epsfig{file=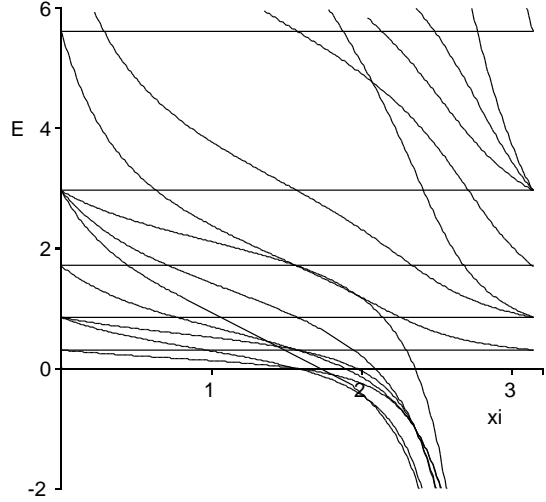,width=8cm}
\end{center}
\caption{\label{k_delta} Lowest energy level in terms of $\xi$ for
2 particles and $\delta$ impurity}
\end{figure}

Let us move on to the case of two particles. The impurity Bethe
equations read
\begin{equation}
\begin{cases}
\exp(2i k_1 L)\, \frac{k_1+k_2-ig}{k_1+k_2+ig}\frac{k_1-k_2-ig}{k_1-k_2+ig}=\lambda_1(k_1)\\
\exp(2i k_2 L)\,
\frac{k_2+k_1-ig}{k_2+k_1+ig}\frac{k_2-k_1-ig}{k_2-k_1+ig}=\lambda_2(k_2)
\end{cases}
\end{equation}
where each of the eigenvalues $\lambda_1(k)$, $\lambda_2(k)$ can
be either $\pm 1$ or $\pm \frac{k\tan \xi +i}{k\tan \xi-i}$. We
display on Figure \ref{k_delta} the corresponding lowest energy
levels.
When the two eigenvalues are $\pm 1$, we obtain the constant
energy levels. Otherwise, when one at least of the eigenvalues is
$\pm \frac{k\tan \xi +i}{k\tan \xi-i}$, the energy levels are
decreasing functions of $\xi$. Again, for special values of $\xi$
($\xi=0$, $\xi=\pi/2$), there are degeneracies which are lifted
when we tune the impurity. Finally, for $\xi> \pi/2$, the lowest
energy levels give rise to bound states with the impurity (we
recall that $\xi\to \pi$ corresponds to $\eta\to -\infty$ \ie an
infinitely negative coupling constant).

\subsection{Asymmetric impurity in a box}

In this paragraph, we give an example of an asymmetric impurity
\ie an impurity which breaks parity invariance. We simply present
the one-particle case for the box boundary conditions of section
\ref{box}. For convenience, we use the parametrization
(\ref{standard}) and the impurity is characterized by a single
parameter as follows
\begin{equation}
\alpha=1~~,~~a=\sin^2 \text{w}~~,~~b=-\cos \text{w}~~,~~c=\cos
\text{w}~~,~~d=1~~,~~\text{w}\in[0,\pi)\,.
\end{equation}

\begin{figure}[htp]
\begin{minipage}[b]{8cm}
 \epsfig{file=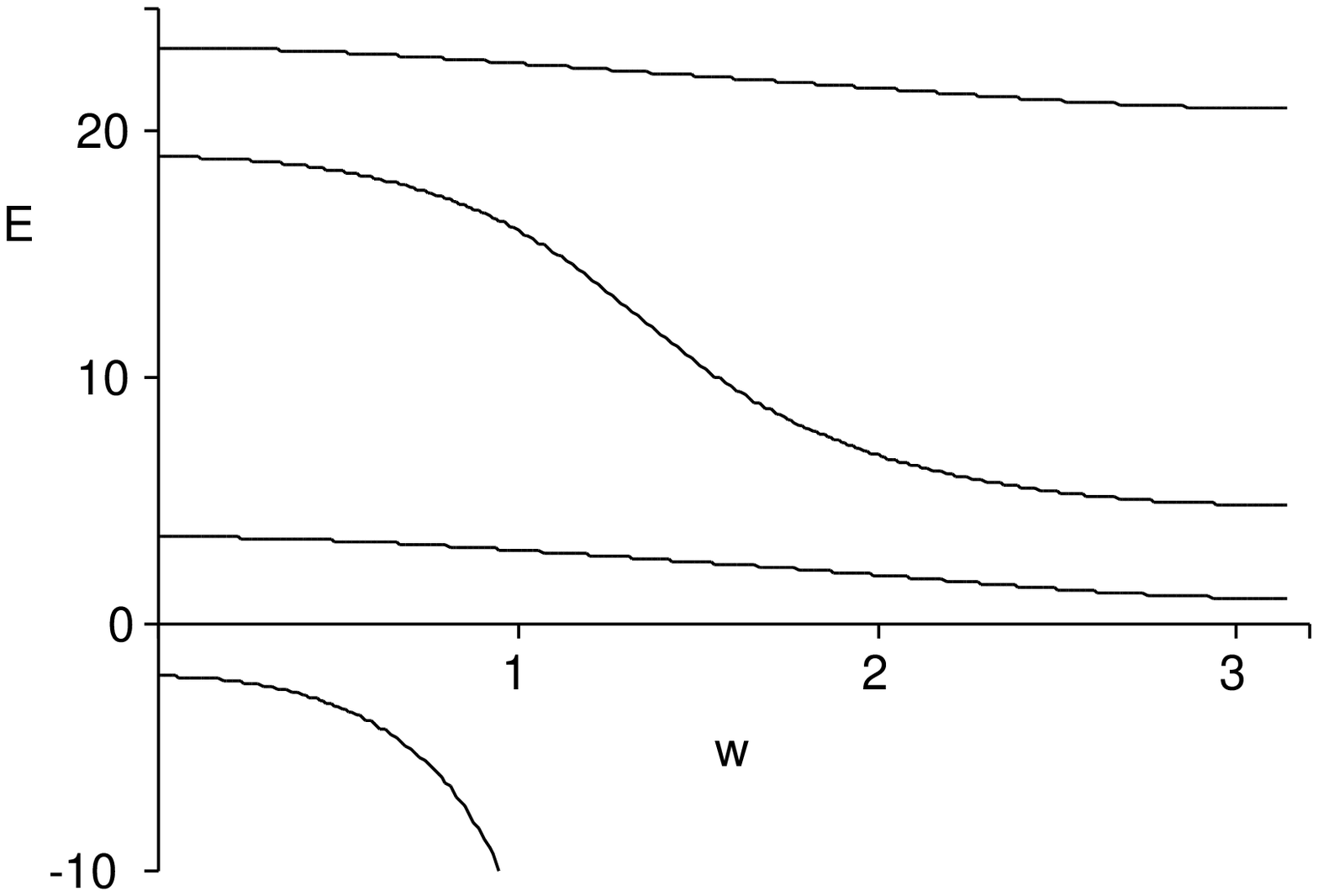,width=8cm}
\caption{\label{energy_gaudin}Lowest energy levels for a parity
breaking impurity}
\end{minipage}
\quad
\begin{minipage}[b]{8cm}
 \epsfig{file=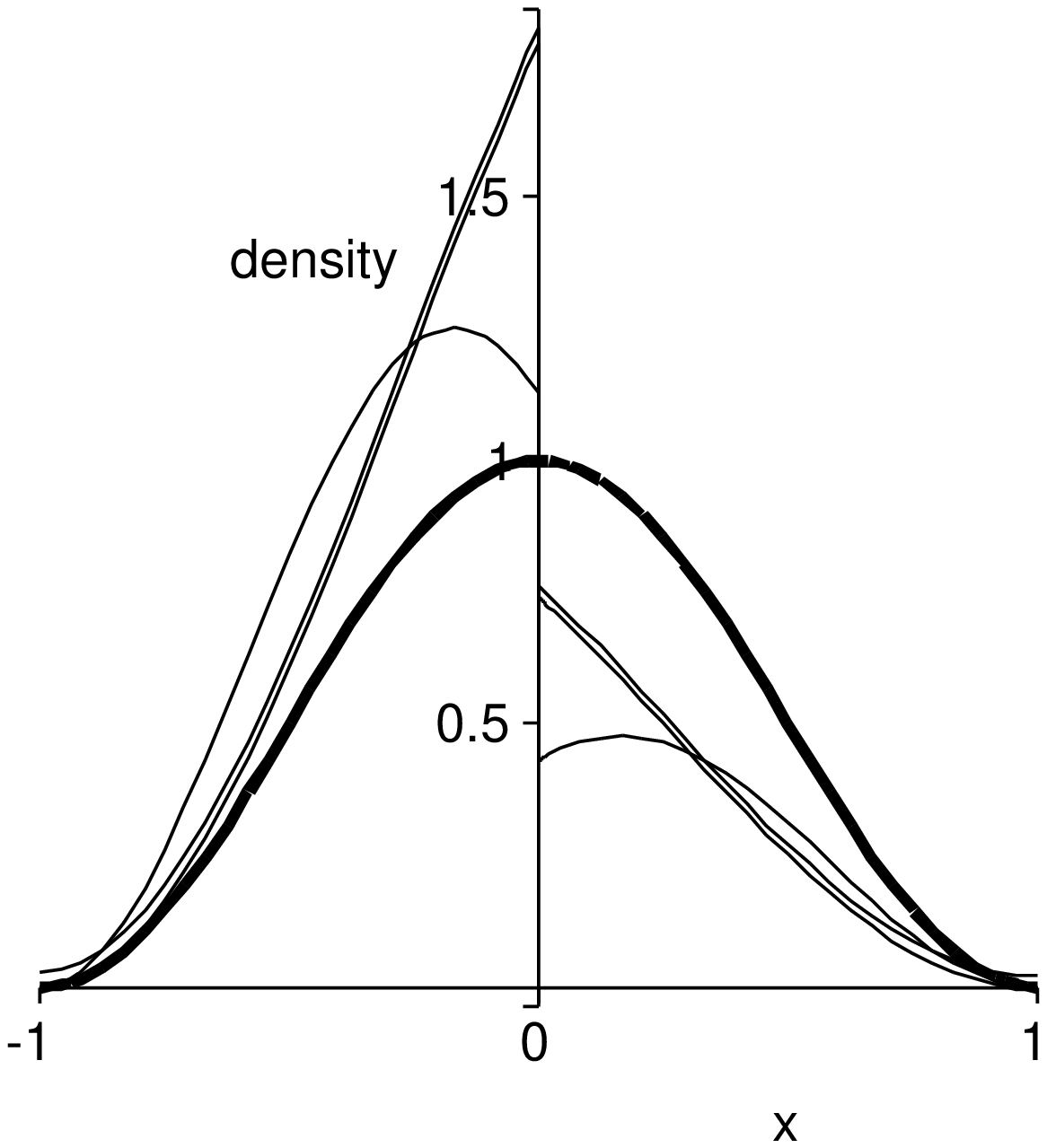,width=8cm}
 \caption{\label{density_gaudin}Density (=$|\varphi(x)|^2$) for the first
positive energy level}
\end{minipage}
\end{figure}

Then the Bethe equations take the form
\begin{eqnarray}
e^{2ikL}=-\frac{(k^2-1)\cos
\text{w}+ik\sqrt{4+\cos^4\text{w}}}{(k^2+1)\cos
\text{w}-ik(\cos^2\text{w}-2)}
\qmbox{or}e^{2ikL}=-\frac{(k^2-1)\cos
\text{w}-ik\sqrt{4+\cos^4\text{w}}}{(k^2+1)\cos
\text{w}-ik(\cos^2\text{w}-2)}
\end{eqnarray}
The two equations are never equivalent so that we do not observe
level crossing (see Figure \ref{energy_gaudin}).
On Figure \ref{density_gaudin}, the density profile for the lowest
positive energy level shows the striking feature of this impurity
for different values of $\text{w}$. One can clearly see the parity
invariance breaking on the thin and double-solid curves
($\text{w}=0$ and $11\pi/12$ respectively). Again for the
particular value $\text{w}=\pi/2$ (the thick curve), the impurity
becomes reflectionless with trivial transmission equal to $1$.
Parity is restored and we observe a single excited mode in a box.

\newpage
\section{Conclusions and outlook}
In this paper, we presented and solved the one-dimensional problem
of $N$ interacting particles in the presence of an impurity. In
the process of the Bethe ansatz for the wavefunction, doubling the
dimension of the representation of the underlying Weyl group was a
crucial ingredient with respect to previous approaches. This is
reminiscent of the general RT algebras framework recently
introduced and allows for an exact treatment of impurities.

We also established the impurity Bethe equations controlling the
energy spectrum. Although some basic understanding emerged from
the study of the simple one and two-particle cases, a systematic
study of the finite size Bethe equations as well as of the
integral equations arising in the thermodynamic limit is needed.
This will give important non-perturbative information on the
effect of the impurity on the system. This should be compared to
the perturbative and effective approach of Kane and Ficher
\cite{KF}. A way to tackle this problem is to compute the sound
velocity $u$ and the Luttinger liquid parameter $K$ directly from
our impurity Bethe equations. The result will actually give the
analogs of the renormalized $u^*$ and $K^*$ in Kane and Fisher's
approach. This issue as well as other physical consequences
deserve careful attention and will be investigated elsewhere.

In any case, we already observed that a tunable impurity can lift
degeneracies in the energy. It can also confine asymmetrically
stationary states. In this respect, we emphasize that the present
approach allows for the description of unusual asymmetric impurities
(and not only the standard "delta impurity") whose effects for finite
size systems and in the thermodynamic limit will also be addressed elsewhere.\\

\vspace{1cm}

\textbf{Acknowledgements:} V.C. thanks the UK Engineering and
Physical Sciences Research council for a Research Fellowship. N.C.
is supported by the TMR Network "EUCLID. Integrable models and
applications: from strings to condensed matter", contract number
HPRN-CT-2002-00325. We acknowledge the warm support of M. Mintchev
and E. Ragoucy.

\end{document}